\documentclass[a4paper,fleqn,usenatbib]{mnras}

\usepackage[T1]{fontenc}
\usepackage{ae,aecompl}

\newcommand{\lsp}{LS~I~+61$^{\circ}$303}
\newcommand{\lsi}{LS~I~+61$^{\circ}$303~}

\newcommand{\psr}{PSR B1259-63~}
\newcommand{\psp}{PSR B1259-63}

\newcommand{\swf}{Swift J1357.2-0933~}
\newcommand{\swp}{Swift J1357.2-0933}
\newcommand{\vcy}{V404 Cygni~}
\newcommand{\vcp}{V404 Cygni}
\newcommand{\beq}{\begin{equation}}
\newcommand{\eneq}{\end{equation}}
\usepackage{graphicx}	
\usepackage{amsmath}	
\usepackage{amssymb}	

\title[The Black Hole Candidate \lsi]{The Black Hole Candidate \lsi}

\author[M. Massi et al.]{
M. Massi,$^{1}$\thanks{E-mail: mmassi@mpifr-bonn.mpg.de}
S. Migliari,$^{2,3}$
and M. Chernyakova$^{4,5}$
\\
$^{1}$Max-Planck-Institut f\"ur Radioastronomie, Auf dem H\"ugel 69,
 D-53121 Bonn, Germany\\
$^{2}$ESAC/ESA, Camino Bajo del Castillo s/n, Urb. Villafranca del Castillo, 28692, Villanueva
de la Cañada, Madrid, Spain\\
$^{3}$Department of Quantum Physics and Astrophysics \& Institute of Cosmos Sciences,
Univ. of Barcelona, Mart\'i i Franqu\`es 1,\\
 08028 Barcelona, Spain\\
$^{4}$School of Physical Sciences, Dublin City University, Dublin 9, Ireland\\
$^{5}$Dublin Institute for Advanced Studies, 31 Fitzwilliam Place, Dublin 2, Ireland
}

\date{Accepted XXX. Received YYY; in original form ZZZ}

\pubyear{2016}

\begin{document}
\label{firstpage}
\pagerange{\pageref{firstpage}--\pageref{lastpage}}
\maketitle

\begin{abstract}
In recent years, fundamental relationships
for the  black hole X-ray binaries  
have  been established
between  
their X-ray  luminosity $L_X$ and the  photon index $\Gamma$
of their X-ray spectrum.
For the moderate-luminosity regime, an anti-correlation between $\Gamma$ and $L_X$  has been  observed.
In this article,  aimed to
verify if the moderate luminous X-ray binary system  \lsi  is  a black hole,
we analyse   {\it Swift} observations  of \lsp. We  compare 
the derived   $L_X$ vs $\Gamma$  distribution, first with the statistical
trend for  black hole X-ray binaries, then
with the trend  of the pulsar \psp, and finally with
the individual trends  of the black hole X-ray binaries \swf and \vcp.
We find that the system \psr   shows   a positive correlation between $\Gamma$ and $L_X$, 
whereas in contrast \lsi shows the same anti-correlation as for 
black hole X-ray binaries.
Moreover, the trend of \lsi in the $L_X/L_{Eddington} - \Gamma$ plane overlaps
with that of the two  black holes \swf and \vcp. 
All three systems, \swp,  \vcy and \lsi  well trace the last part of the evolution of accreting black holes at  
 moderate-luminosity  until their drop to  quiescence.
\end{abstract}


\section{Introduction}
 X-ray binaries are stellar systems formed by a normal stellar component
 and a compact object 
that could be a black hole or a neutron star \citep{lewin97}.
The usual method  for distinguishing the nature of the compact object
is to determine its mass. 
In fact, 
\citet{kalogerabaym96} determined that the maximum 
gravitational mass of a neutron star for stability 
against gravitational collapse is $2.2-2.9 M_{\odot}$,
with 2.9 $M_{\odot}$ as the safest estimate.
Therefore, a compact object with  mass  above 2.9 $M_{\odot}$,
is identified as black hole candidate \citep{kalogerabaym96}.
The mass  is determined by the mass function obtained by optical observations \citep{lewin97}:
\beq
f ={M_{\rm X}^3 sin^3 i \over (M_{\rm X} +M)^2}
\eneq
where, $M_X$ is  the mass of the compact object,
$M$  is the mass of the companion star, and  $i$ is the inclination of the orbital plane.
Another possible way to determine a black hole in a X-ray system results  directly from
X-ray observations.
Indeed, there is a  clear relationship between the photon index $\Gamma$ and the X-ray luminosity $L_X$
for black hole X-ray binaries.
The latest statistical study  of black hole X-ray  binaries \citep{yang15a, yang15b}  
shows that over a luminosity interval covering  more than three orders of magnitude,  
$\sim 10^{33} \leq  L_X \leq 10^{36.5}$ erg/s, there is an anti-correlation between $\Gamma$ and 
the 2-10 keV luminosity, $L_X$,
in the form of 
$\Gamma=(-0.11\pm0.01){\rm log}_{10}L_X+(5.63\pm0.017)$.

The system \lsi is
one of the very few radio emitting X-ray binaries  being
 associated with a source of very high energy gamma ray emission  \citep{albert06}.
In this X-ray  binary,   the compact object
   orbits around a Be star with an 
    orbital period equal to $26.4960 \pm 0.0028$ d
    \citep[][]{hutchingscrampton81, gregory02}.
As described in Sec. 2, the mass function does not allow a precise determination
of the mass of the compact object in \lsp, and consequently is still under debate 
if  a neutron star or a black hole is the engine of 
this rare X-ray binary emitting all over the
electromagnetic spectrum.
In the present work, aimed to  discriminate the nature of the compact object in \lsp,
we investigate if the X-ray characteristics of \lsi fit with those of  black hole X-ray  binaries.
We present our analysis of  {\it Swift} observations  of \lsi (Sect. 3) and compare 
the derived   $L_X$ vs $\Gamma$  distribution with 
the trend  of the pulsar \psr and with 
the trend  of the black holes  (Sect. 4).
Our conclusions are presented in Sect. 5.

\section{The  microquasar and pulsar scenarios}
 A small percentage of X-ray binaries, 
called microquasars,
 have  radio emitting jets,
probed either directly by  high resolution radio images,
or indirectly by their radio spectrum, that, for a steady jet is flat
\citep{spencer96, mirabelrodriguez99, fender01}.
    In a microquasar the radio jet is related to the  behaviour of
an  accretion disk around  a black hole,
as in High Mass and Low Mass X-ray
binary systems (HMXB, LMXB) or to an accretion disk
around a neutron star with a low ($\leq 10^{8}$ G)  magnetic field as
in LMXB \citep{migliarifender06}.
Besides from accretion-powered radio jets, 
radio emission 
is observed also from X-ray systems, such as 
\psp,  having as compact object  a fast rotating, non accreting neutron star
with strong magnetic-field ($\sim 10^{12}$ G).
In this case the
observed continuum radio emission 
is  due to particles  accelerated at the
 shock  between the  relativistic wind of the pulsar and the wind of the companion star
\citep{tavani94, moldon12, chernyakova14}.

The strong  radio emitting HMXB  \lsp,
 proposed to host a black hole  
\citep{punsly99},
  shows  resolved  radio jets
\citep[][and reference therein]{massi12}
and a flat radio spectrum \citep{zimmermann15}.
In particular, the fact that the flattening of the spectrum occurs twice along the
orbit suggests two main accretion events along the orbit \citep{massikaufman09,
jaronetal16}. 
Specifically,  \citet{kaufman02} suggested  \lsi be
   a precessing microblazar.
In this kind of microquasars,
the   precession  brings the approaching jet component closer to the line of sight
and the approaching jet component gets boosted. On the contrary,  the receding jet component is de-boosted,
an effect observed  in \lsi with  a sequence of  VLBA images
showing the change from a two-sided jet to an one-sided structure
\citep[fig. 7]{massitorricelli14}.
In contrast to the microquasar scenario, 
the one-sided radio jet of \lsi was alternatively interpretated by
\citet{dhawan06} as the cometary tail predicted by \citet{Dubus2006}
for the outflowing shocked wind material resulting from the interaction
between a  pulsar wind and the companion wind.
The hypothesis of  a strong magnetic field neutron star 
in \lsi was considered   also by \citet{torres12}  
assuming an association of the system to a magnetar event 
detected in a crowded X-ray field with other potential candidates besides
\lsi
\citep{reatorres08}.

\begin{figure}
\includegraphics[width=0.42\textwidth,angle=0]{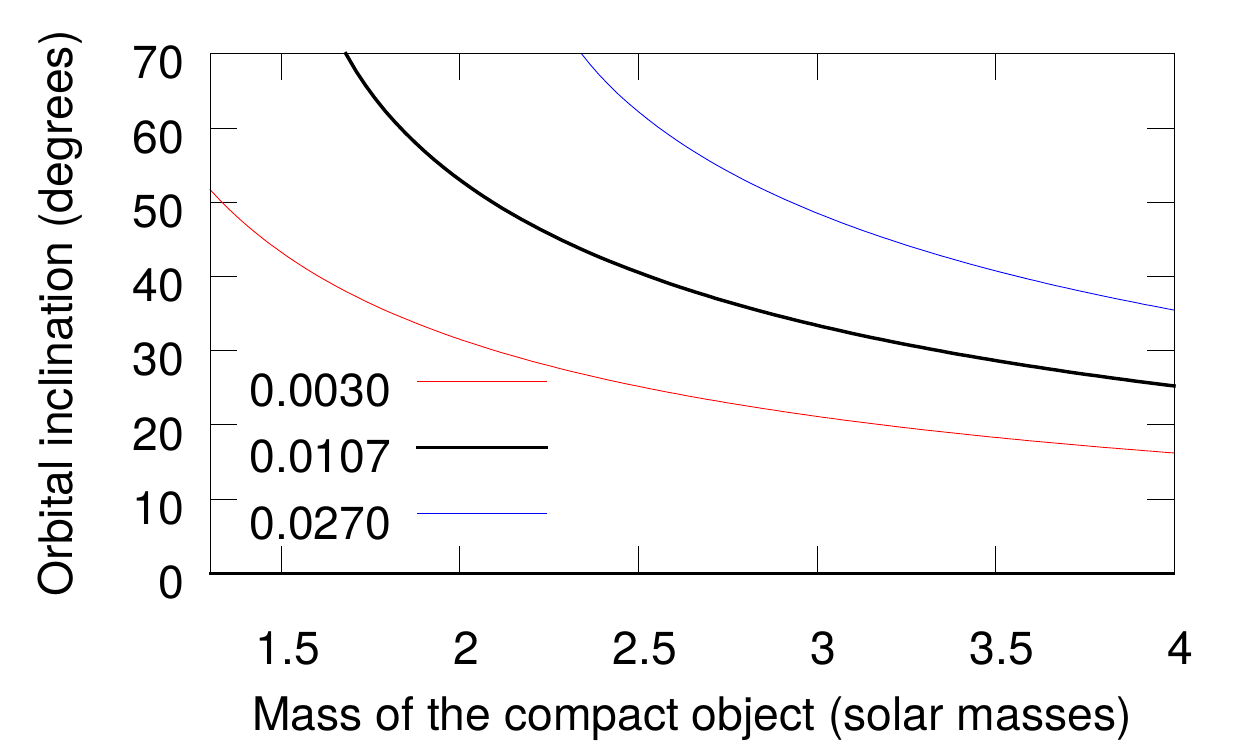}
\caption{Mass of the compact object in \lsi 
as a function of the inclination angle,
 for three different values of the  mass function    
 \citep{casares05} (see Sec. 2).} 
\label{fig:massf}
\end{figure}

\citet{casares05} used  optical spectroscopy to measure a mass function  $f=0.0107~M_{\odot}$  of \lsp.
     Optical polarization observations  have determined
a value of $\sim$25 degrees for the
     rotational axis, $r$,  of the Be disk in \lsi \citep{nagae06}.
     For parallel  orbital and Be spin axes, i.e., $i=r$,
and a mass of the B0 star $M=17.5~M_{\odot}$   \citep{townsend04},
the mass of the compact object in \lsi is equal to 4 $M_{\odot}$ (Fig. 1),
clearly above the  2.9 $M_{\odot}$ limit of 
\citet{kalogerabaym96} and therefore
indicates a black hole.
However,  the range of uncertainty of the mass function is rather large, $0.003-0.027~M_{\odot}$ (Fig. 1), and
in addition the $i$ and $r$  axes could be tilted.
This means that optical observations  only give a hint for the presence of a black hole
in \lsp.  In the next sections  we analyse  if this hypothesis is  
corroborated or ruled out by X-ray observations.

     \begin{figure}
	     \center
	     \includegraphics[width=7.0cm,angle=0]{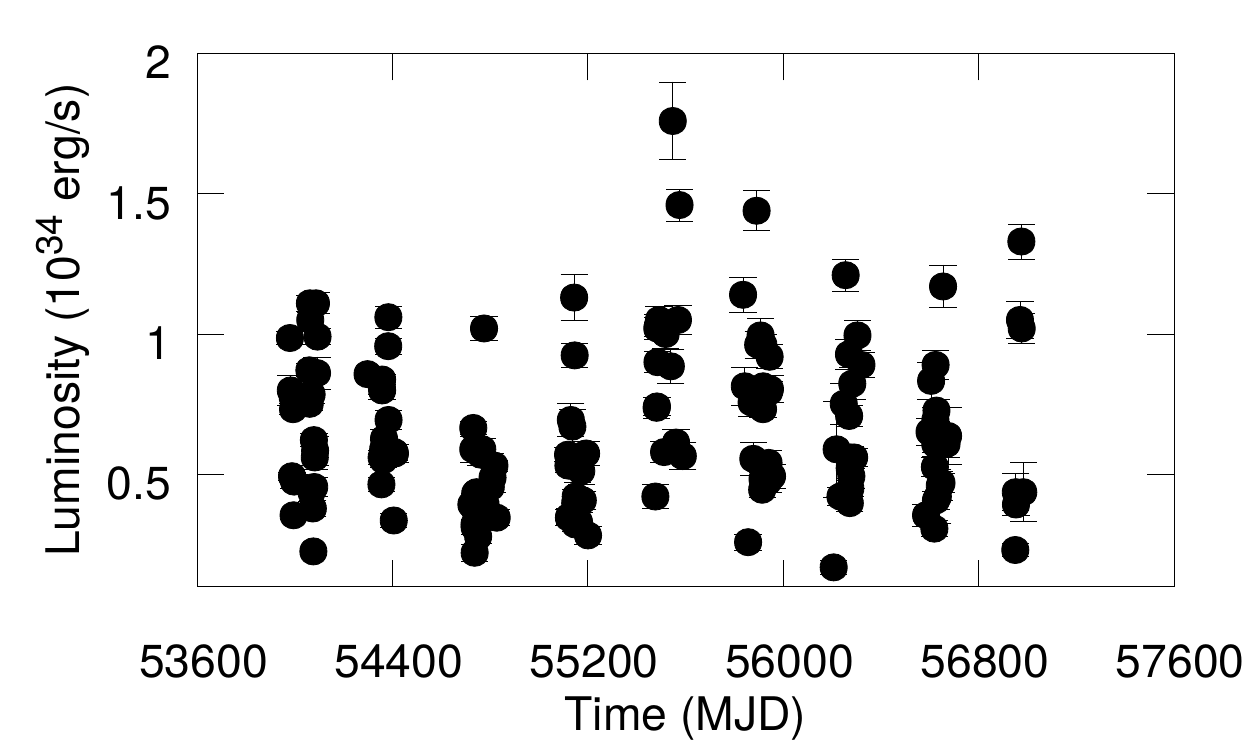}\\
\caption{X-ray light curve (1-10 keV) of {\it Swift} \lsi observations.} 
\end{figure}

\section{Data}
We  analyse {\it Swift}/XRT   (1-10 keV) data in the interval 53980$-$56983 MJD 
using  {\tt FTOOLS/HEASOFT 6.14} tools. 
\lsi was observed both in Photon Counting (PC) and Windowed Timing (WT)
 modes. The
 initial cleaning of events is done using {\tt xrtpipeline} with
 standard parameters (http://www.swift.ac.uk/analysis/xrt/index.php). 
For the source extraction region in the PC mode
 we use a
 circle with radii from 5 to 30 pixels
 depending on the count rate \citep{evans09}; for the WT mode the radius
 of the
 source extraction region is selected to be equal to 25 pixels. To
 collect the
 background we use an
 annulus region with an inner (outer) radius of 60 (110) pixels in both
 observational modes. The count rate from the source was too low to
 pile up the
 detector in any of the  observations.
Selection of data points with better  than 3$\sigma$ detection  
        in both flux and photon index 
results in 150 {\it Swift}   data points shown in Fig. 2.

We also take
into account published historical data  with the better than $3 \sigma$ 
detection level: 
{\it XMM-Newton} data for \lsi 
\citep[recomputed at 1-10 keV]{sidoli06, anderhub09};    
 {\it Swift}  and {\it XMM-Newton} 
data for the pulsar \psr 
 \citep{chernyakova06, chernyakova09,chernyakova14,chernyakova15};
{\it Swift} 
data for the black hole X-ray binary \swf \citep{armaspadilla13} and
{\it Chandra} 
data for the black hole X-ray binary \vcy \citep{plotkin17}. 
For the  luminosity calculation we  assume a $2.0\pm0.2$ kpc distance to \lsi   \citep{frailhjellming91},
2.3 kpc distance to \psr  \citep{negueruela11}, 
2.29 kpc distance to \swf \citep{matasanchez15},  $2.39 \pm 0.14$ kpc distance to \vcy \citep{millerjones09}.

\section{Results and Discussion}
In this section we present  
our results for \lsi and  \psr  and compare them 
with the results of \citet{yang15b}
for  a statistical sample of black hole X-ray binaries. 
We also compare here the behaviour of  \lsi
with that of the two moderate-luminosity black hole X-ray binaries \vcy and \swp.

In the left panel of Fig. 3  {\it Swift} data (red circles) and {\it XMM-Newton} data (red squares)
of \lsi in the $L_X - \Gamma$ plane are presented . The data cover the luminosity range
$2.0 \times 10^{33}\leq L_X \leq 1.8 \times 10^{34}$ erg/sec.
The data from the two different satellites 
(averaged over luminosity bins of  $0.6 \times 10^{33}$ erg/sec)
are given separately to show 
the  agreement of  the two different data sets:
points at the same luminosity result in the same photon index.
The combined and  averaged  
 over luminosity bins of  $0.6 \times 10^{33}$ erg/sec data from both satellites, are given in Table 1.
The fit of these averaged \lsi data  results  in:
\beq
     \Gamma_{\rm{LS~I~+61303}}=(-0.13\pm0.09){\rm log}_{10}L_X+(6\pm3)
\eneq
with reduced $\chi^2$=1.2.
The fit is shown with a red line  in the left panel of Fig. 3.
We compare this fit of \lsi with   
that  of the statistical sample of
black hole X-ray binaries at   moderate-luminosity, i.e.,
$\sim 10^{33} \leq L_X \leq 10^{36.5}$ erg/sec
by  \citet{yang15b}:
\beq
\Gamma_{\rm{Black~Holes}}=(-0.11\pm0.01){\rm log}_{10}L_X+(5.63\pm0.017)
\eneq
Clearly the two independent 
fits determine the same slope of $\sim -0.1$ for \lsi and the statistical sample of black hole X-ray binaries.
The black line drawn in the left panel of Fig. 3 corresponds to $\Gamma=-0.12~{\rm log}_{10} L_X + 5.63$.
We perform now the same comparison for the non accreting pulsar \psp.
The right panel  Fig. 3  show  
 {\it Swift} data (blue  circles) and {\it XMM-Newton} data (blue squares).
Our least square fit for averaged  \psr data  
results in:
\beq
\Gamma_{\rm{PSR~B1259-63}}=(+0.47 \pm 0.05)~{\rm log}_{10} L_X - (15 \pm 2)
\eneq
with reduced $\chi^2$=19. In  Fig. 3 the fit for \psr is represented by a  blue line.
The  slope, in this case is $+0.47 \pm 0.05$. That is in the considered luminosity
range an increasing  
luminosity corresponds to  an increasing 
 photon index. This is clearly different from
the  anti-correlation between $\Gamma$ and luminosity   typical for  
black hole X-ray binaries in the same luminosity range.
\psr data cover the   luminosity range of
 $2.0 \times 10^{32}\leq L_X \leq 3.4 \times 10^{34}$ erg/sec
This is a bit larger than that covered by \lsp.
Restricting the  linear fit for \psr data 
within the  same luminosity interval of \lsi,
i.e.,  $2.0 \times 10^{33}\leq L_X \leq 1.8 \times 10^{34}$ erg/sec,
does not change the result for \psr of a positive correlation between 
luminosity and photon index, the fit in fact results in: 
$\Gamma=(+0.37 \pm 0.08)~{\rm log}_{10} L_X - (11 \pm 3)$
 with reduced $\chi^2$=22.

We compare in the top panel of  Fig. 4 \lsi with the two black hole
binaries \swf and \vcp.
The mass of the black hole in \vcy is of $9.0^{+0.2}_{-0.6} M\odot$ 
 \citep[][and references therein]{plotkin17}; the mass of the black hole in \swf is 
$12.4\pm 3.6 M\odot$ \citep{casares16}.
From the optical observations of \lsi discussed in the introduction (Fig. 1) the mass of
\lsi is determined to be about 4 $M\odot$.
The comparison of black hole X-ray binaries with different masses
is  done (e.g., \citet{yang15a}) scaling the luminosity by the 
 Eddington Luminosity, 
$L_{Eddington} = 1.26 \times  10^{38}$ [$M_X$/M$_{\odot}$] erg/sec. 
In the bottom panel of Fig. 4 we show  the three systems, \lsp, \vcy and \swf 
in the $L_X/L_{Eddington} - \Gamma$ plane.
Besides  one point at $L_X/L_{Eddington}= 3.5 \times 10^{-5}$  where
 the photon index of  \lsi deviates  more than one sigma from the photon index value
of \swf at the same luminosity, all other  
\lsi data line up and overlap with  \swp.
At $L_X/L_{Eddington}= 0.9 \times 10^{-5}$, all three systems,
\lsp, \swf and \vcp, overlap.
The  overlap continues between \vcy and \lsi and 
 at the lowest  luminosities, where
the prediction \citep{plotkin13, yang15a,yang15b}
for black hole binaries dropping
in their quiescent state 
is a photon index $\Gamma\sim 2$, 
all three systems, \lsp, \swf and \vcy  converge on that value.
\begin{figure*}[t]
	\center
\includegraphics[width=8.0cm,angle=0]{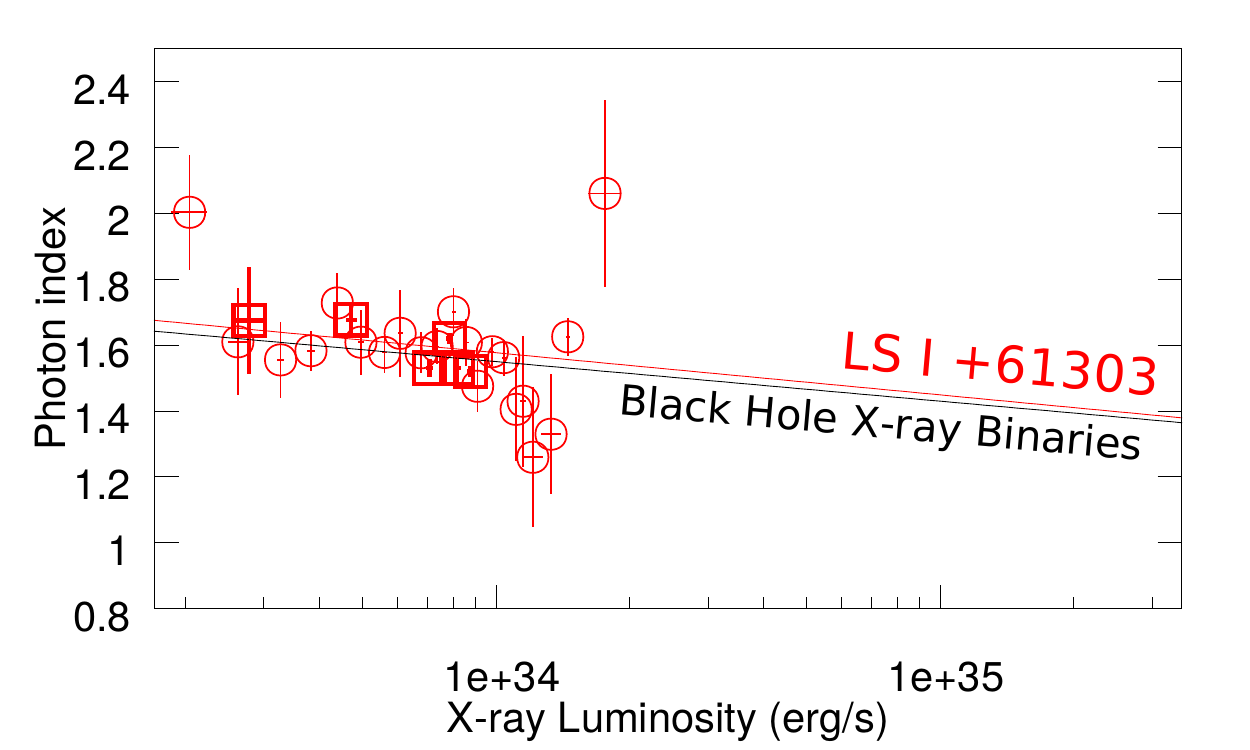}
\includegraphics[width=8.0cm,angle=0]{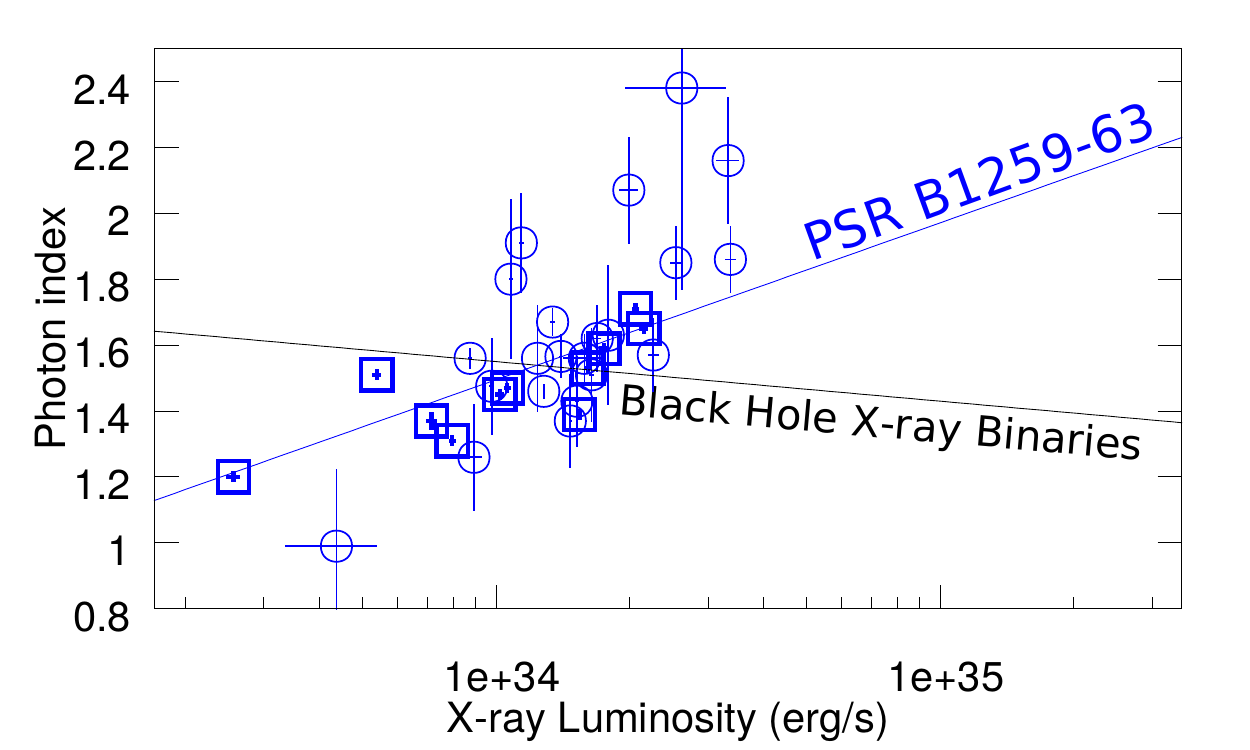}\\
\caption{$L_X-\Gamma$ plane: \lsi (red) and  \psr (blue) data with their fits.
Circles indicate {\it Swift} data,  
and squares {\it XMM-Newton} data.
The black  line is the equation for black hole X-ray binaries 
 $\Gamma= - 0.12~L_X+5.63$ (see Sect. 4).
}
\end{figure*}
\begin{figure*}[t]
\includegraphics[width=9cm,angle=0]{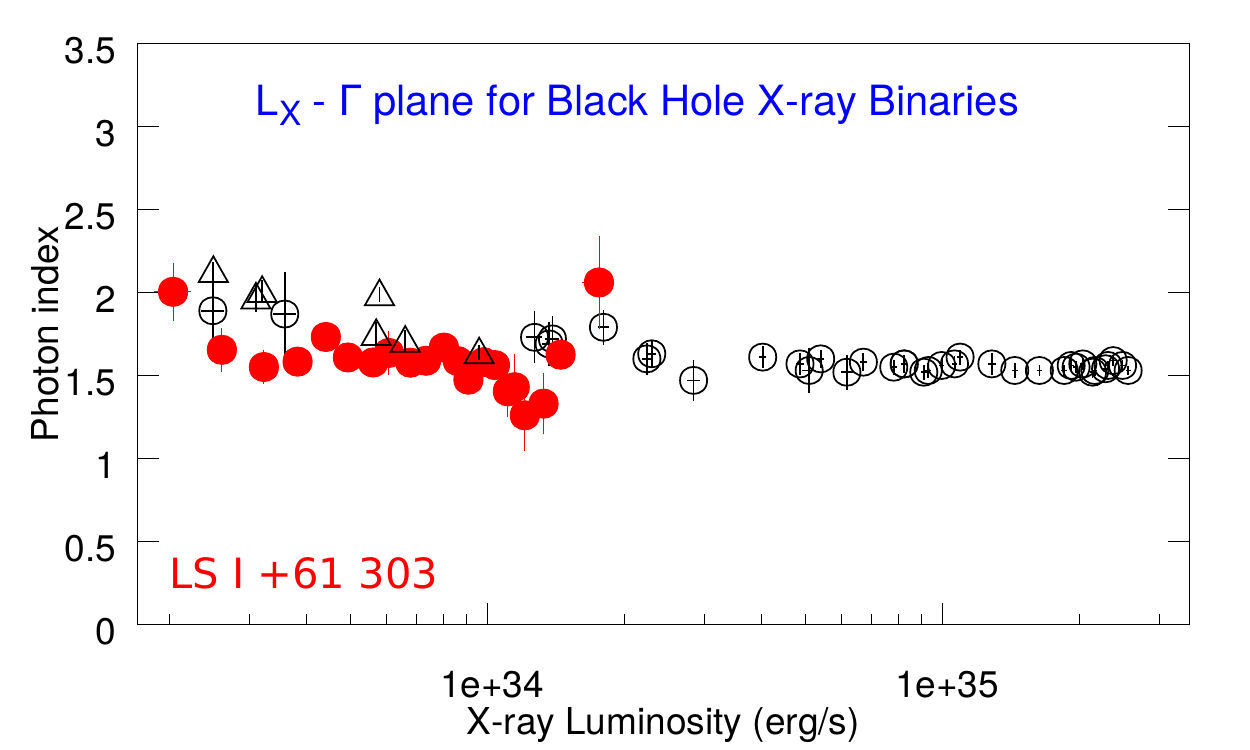}\\
\includegraphics[width=14cm,angle=0]{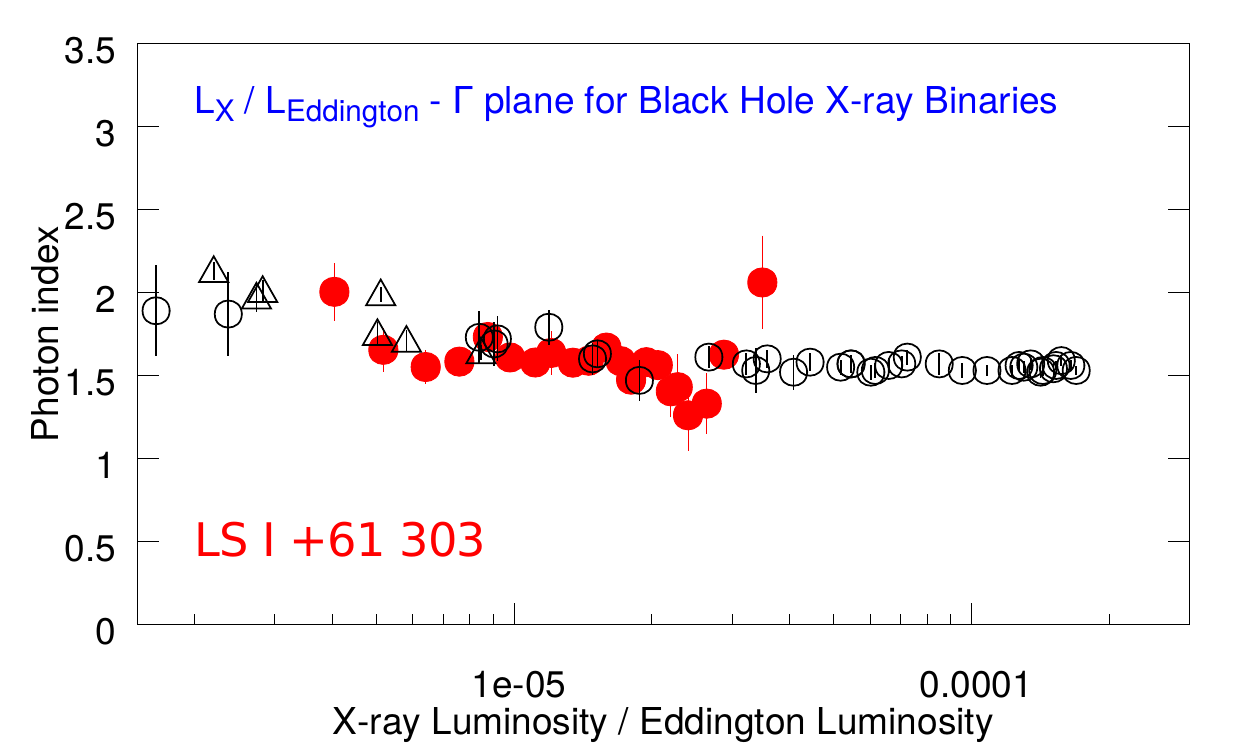}\\
\caption{
\lsi (red circles) and the black hole binaries \vcy (black triangles) and \swf (black circles)
 in the $L_X-\Gamma$ plane (top) and in the ${L_X/L_{Eddington}}-\Gamma$ plane (bottom).
The \lsi data are those of Tab.1. The assumed masses are: 4$M\odot$ for \lsp, 9$M\odot$ for \vcy
and 12 $M\odot$ for \swf (see Sec.4).
}
\end{figure*}
 
\begin{table}
	\centering
	\caption{ Photon index and luminosity of \lsi as observed by  {\it Swift} and {\it XMM-Newton}. Data are combined and averaged over luminosity bins of 0.6$\times 10^{33}$ erg/s.}
	\label{tab:example_table}
	\begin{tabular}{lccr} 
		\hline
		$\Gamma$ & $\Delta \Gamma$ & $L_X$ & $\Delta L_X$\\
		         &                 & ($10^{33}$ erg/s)&  ($10^{33}$ erg/s)\\
		\hline
%
     2.00 &     0.17 &     2.04 &     0.18 \\
     1.65 &     0.13 &     2.61 &     0.09 \\
     1.55 &     0.10 &     3.23 &     0.06 \\
     1.58 &     0.06 &     3.83 &     0.06 \\
     1.73 &     0.07 &     4.42 &     0.04 \\
     1.61 &     0.08 &     4.94 &     0.04 \\
     1.58 &     0.06 &     5.61 &     0.04 \\
     1.64 &     0.13 &     6.08 &     0.05 \\
     1.58 &     0.06 &     6.78 &     0.08 \\
     1.59 &     0.05 &     7.34 &     0.05 \\
     1.67 &     0.06 &     8.03 &     0.04 \\
     1.59 &     0.05 &     8.62 &     0.07 \\
     1.48 &     0.07 &     9.09 &     0.07 \\
     1.58 &     0.04 &     9.81 &     0.06 \\
     1.56 &     0.05 &    10.40 &     0.05 \\
     1.41 &     0.15 &    11.10 &     0.00 \\
     1.43 &     0.20 &    11.50 &     0.12 \\
     1.26 &     0.21 &    12.10 &     0.57 \\
     1.33 &     0.18 &    13.30 &     0.62 \\
     1.62 &     0.06 &    14.50 &     0.10 \\
     2.06 &     0.28 &    17.60 &     1.39 \\
		\hline
	\end{tabular}
\end{table}

\section{SUMMARY AND CONCLUSIONS}

High resolution images show a morphological change  of the radio structure 
of the system \lsp. 
Two scenarios have been proposed:
 the black hole - microquasar/microblazar scenario and
the neutron star - pulsar/magnetar scenario.
Optical observations indicate that the compact object in \lsi could be a black hole,
but a neutron star cannot be excluded.
In this work, as direct  test of  the  black hole hypothesis   
we compare the X-ray characteristics of \lsi with
those of black hole X-ray binaries in the same luminosity range.
Our results are:
\begin{enumerate} 
\item
the trend  in the $L_X - \Gamma$ plane
of \lsi is very different from that of the non-accreting pulsar \psr.
This result is at odds with the hypothesis \citep[e.g.,][]{moldon12b} 
that the  emission from the two systems could be  due to the same
physical process, i.e., the collision between a pulsar wind and the outflow of the
companion star. 
\item
the trend of \lsi in the  $L_X$ -- $\Gamma$ plane is in a good agreement  with
that of moderate-luminosity regime  black holes in general and with  \swf 
and \vcy 
in particular. 
This result corroborates the black hole - microquasar scenario for \lsi \citep[e.g.,][]{massi04}
where the emission is due to the accretion-ejection physical processes.
\end{enumerate} 
        These results point towards   \lsi as the second known binary 
system where a    
        black hole is orbiting a fast rotating B star,
 after the  case of MWC 656 \citep{casares14}. 

\section*{Acknowledgements}
 We would like to  thank  
J\"urgen Neidh\"ofer,
 Alessandro Ridolfi and Richa Sharma
 for interesting discussions.
 S.M. acknowledges support by the Spanish Ministerio de Econom\'ia y Competitividad (MINECO) under grants AYA2013- 47447-C3-1-P, MDM-2014-0369 of ICCUB (Unidad de Excelencia 'Mar\'ia de Maeztu')


%
\bsp	
\label{lastpage}
\end{document}